\documentclass{emulateapj}
\usepackage{graphicx}

\shorttitle{THE FORMATION OF RELATIVISTIC BINARY PULSARS LIKE PSR J0737-3039  }
\shortauthors{B. Willems and V.\ Kalogera}

\begin{document}

\title{Constraints on the Formation of PSR\,J0737-3039: the most probable
isotropic kick magnitude
} 

\author{B. Willems and V. Kalogera}
\affil{Northwestern University, Department of Physics and Astronomy,
  2145 Sheridan Road, Evanston, IL 60208, USA}
\email{b-willems@northwestern.edu, vicky@northwestern.edu}

\submitted{Accepted by ApJ Letters, 2004 January 27}

\begin{abstract}
A strongly relativistic binary pulsar has been recently discovered
with the 64m Parkes telescope (Burgay et al.\ 2003). Here we use the
measured properties of this binary (masses and orbital characteristics
as well as age estimates), and we derive the complete set of
constraints imposed on the physical properties of the binary pulsar
progenitor right before the second supernova explosion. We find that:
(i) according to our current understanding of neutron-star formation,
the helium-rich progenitor of the second neutron star is most likely
overflowing its Roche lobe; (ii) the neutron-star kick magnitude is
constrained in the range 60-1560\,km\,s$^{-1}$, with the most probable
value being equal to 150\,km\,s$^{-1}$. While the first conclusion is
in agreement with Dewi \& van den Heuvel (2003), our upper limit on
the kick magnitude is significantly larger than that derived by these
authors. We find that the difference arises because Dewi \& van den
Heuvel (2003) inadvertently neglected to consider kicks directed out
of the pre-supernova orbital plane.
\end{abstract}


\section{Introduction}

The significance for relativistic astrophysics of close binaries with
two neutron stars, one of which is detected as a recycled pulsar, has
been recognized for many years, since the discovery of the first such
system (Hulse \& Taylor 1975). More than 10 years since the discovery
of the second relativistic double neutron star (DNS) PSR\,B1534+12
(Wolszczan 1991), the discovery of the third system in the disk has
been recently announced with important implications for the current
expectations for the detection of DNS inspiral by ground-based
interferometers (Burgay et al.\ 2003; Kalogera et al.\ 2003). This new
system has broken a number of barriers already: it harbors the {\em
fastest} PSR (spin period of 22\,ms) in all DNS, in the {\em tightest}
orbit (orbital period of 2.4\,hr) with the {\em smallest} eccentricity
(0.09) of all DNS. In addition, Lyne et al. (2004) recently reported
the discovery of a 2.8\,s pulsar companion to the millisecond pulsar,
making this the first observed {\it double pulsar} system. 

A number of earlier studies of DNS binaries (e.g. Fryer \& Kalogera
1997, and references therein) have examined the evolutionary history
of the previously known systems and have derived some constraints
related to the formation of the second neutron stars in the
binaries. In this {\em Letter} we consider the recent evolutionary
history of PSR\,J0737-3039 and derive a set of necessary constraints
on the supernova kick imparted to the second neutron star (NS) and on
its progenitor characteristics. We compare our results to those
obtained very recently by Dewi \& van den Heuvel (2003, hereafter
DvdH) and we comment on the origin of their derived constraints.

\section{Orbital Evolution and Dynamics}

The orbital characteristics of the observed DNS systems are primarily
determined by the effects of the supernova explosion (SN) leading to
the birth of the second NS and by the subsequent loss of orbital
energy and angular momentum via gravitational radiation. In the case
of a symmetric SN explosion, the semi-major axis and eccentricity of
the post-SN orbit are uniquely determined by the amount of {\em mass
lost} from the system during the explosion. If, on the other hand, the
SN explosion is asymmetric as currently thought, the post-SN orbital
parameters also depend on the {\em magnitude and direction} of the
kick imparted to the NS.

In agreement with the current understanding of DNS binaries (Tauris \&
van den Heuvel 2004) and as we will show in what follows, the tight
orbit essentially constrains the pre-SN orbital separation to such
small values that the binary just before the explosion is expected to
consist of a helium star (the progenitor of the second NS) and the
first NS. This pre-SN orbit is expected to be circular due to
Roche-lobe overflow (RLOF) from the helium-star progenitor during
earlier evolutionary phases. We constrain the mass $M_0$ of the
helium star, the pre-SN orbital separation $A_0$, and the magnitude
$V_{\rm k}$ of the kick velocity by considering (i) the orbital
evolution of the DNS due to gravitational radiation, and (ii) the
orbital dynamics of asymmetric SN explosions (see also Fryer \&
Kalogera 1997; DvdH).

We first determine the post-SN semi-major axis $A$ and orbital
eccentricity $e$ from the currently observed values $A_{\rm cur}$ and
$e_{\rm cur}$ by integrating the equations derived by Junker \&
Sch\"{a}fer (1992) for the evolution of the orbit due to gravitational
radiation backwards in time. To this end, we use the values $A_{\rm
cur}=1.26\,R_\odot$ and $e_{\rm cur}=0.0878$ reported by Burgay et
al. (2003). For the masses of the pulsar and its companion, we use the
values $M_{\rm p}=1.34\,M_\odot$ and $M_{\rm c}=1.25\,M_\odot$
expected for an edge-on orbit (see Burgay et al. 2003).  Since the
characteristic age is derived under the assumption of a birth spin
period much smaller than the current spin period, it may be quite
unreliable as an estimate for the true age of a recycled pulsar. The
characteristic age derived for the second (not recycled) pulsar in
PSR\,J0737-3039 is equally uncertain because the spin evolution of the
second-born NS is likely to be affected by torques exerted by its
millisecond pulsar companion (Lyne at al. 2004).  We therefore assume
that the first-born NS was recycled to maximum spin-up for
Eddington-limited accretion and use an estimate for the time
$\tau_{\rm b}$ since the pulsar left the spin-up line as an {\em upper
limit} to the time $T_{\rm SN}$ elapsed since the last SN explosion
(see Arzoumanian et al.\ 1999 for details).  The post-SN orbital
parameters at $T_{\rm SN} =\tau_{\rm b} = 100\,$Myr are calculated to
be $A=1.54$ and $e=0.12$. We note that the orbital evolution due to
gravitational radiation is relatively slow for this system so that the
values of $A$ and $e$ are not greatly sensitive to the adopted value
of $T_{\rm SN}$.

Next, we consider the orbital dynamics of asymmetric, instantaneous SN
explosions. As in the past we use the conservation laws of orbital
energy and angular momentum to relate the pre-SN parameters $\left(
A_0, M_0, M_{\rm p} \right)$ and the kick velocity $\vec{V_{\rm k}}$
to the post-SN parameters $\left( A, e, M_{\rm c}, M_{\rm p} \right)$:
\begin{equation}
V_{\rm k}^2 + V_{\rm r}^2 + 2\, V_{\rm k}\, V_{\rm r}\, \cos \theta 
 = G \left( M_{\rm p} + M_{\rm c} \right) \left( {2 \over A_{\rm 0}} 
 - {1 \over A} \right),  \label{eq1}
\end{equation}
\begin{eqnarray}
A_0^2 \left[ V_{\rm k}^2\, \sin^2 \theta \cos^2 \phi \right. 
 & + & \left. \left( V_{\rm k}\, \cos \theta + V_{\rm r}
 \right)^2 \right]  \nonumber \\
 & = & G \left( M_{\rm p} + M_{\rm c} \right) A 
 \left( 1 - e^2 \right), \hspace{1.2cm} \label{eq2}
\end{eqnarray}
where $G$ is the gravitational constant, and $V_{\rm r} = \left[ G 
\left( M_{\rm p} + M_{\rm 0} \right)/A_0 \right]^{1/2}$ is the
relative orbital velocity of the helium star just before its SN
explosion (e.g., Hills 1983; Kalogera 1996). The angles $\theta$ and
$\phi$ describe the direction of the kick velocity: $\theta \in 
[0,\pi]$ is the polar angle between the kick velocity and the relative
orbital velocity of the helium star just before the SN explosion, and
$\phi \in [0,2\,\pi]$ is the corresponding azimuthal angle defined so
that $\phi=0$ represents a plane perpendicular to the line connecting
the centers of mass of the binary components (see Kalogera 2000
for a graphic representation).
 
\begin{figure*}
\resizebox{8.8cm}{!}{\includegraphics{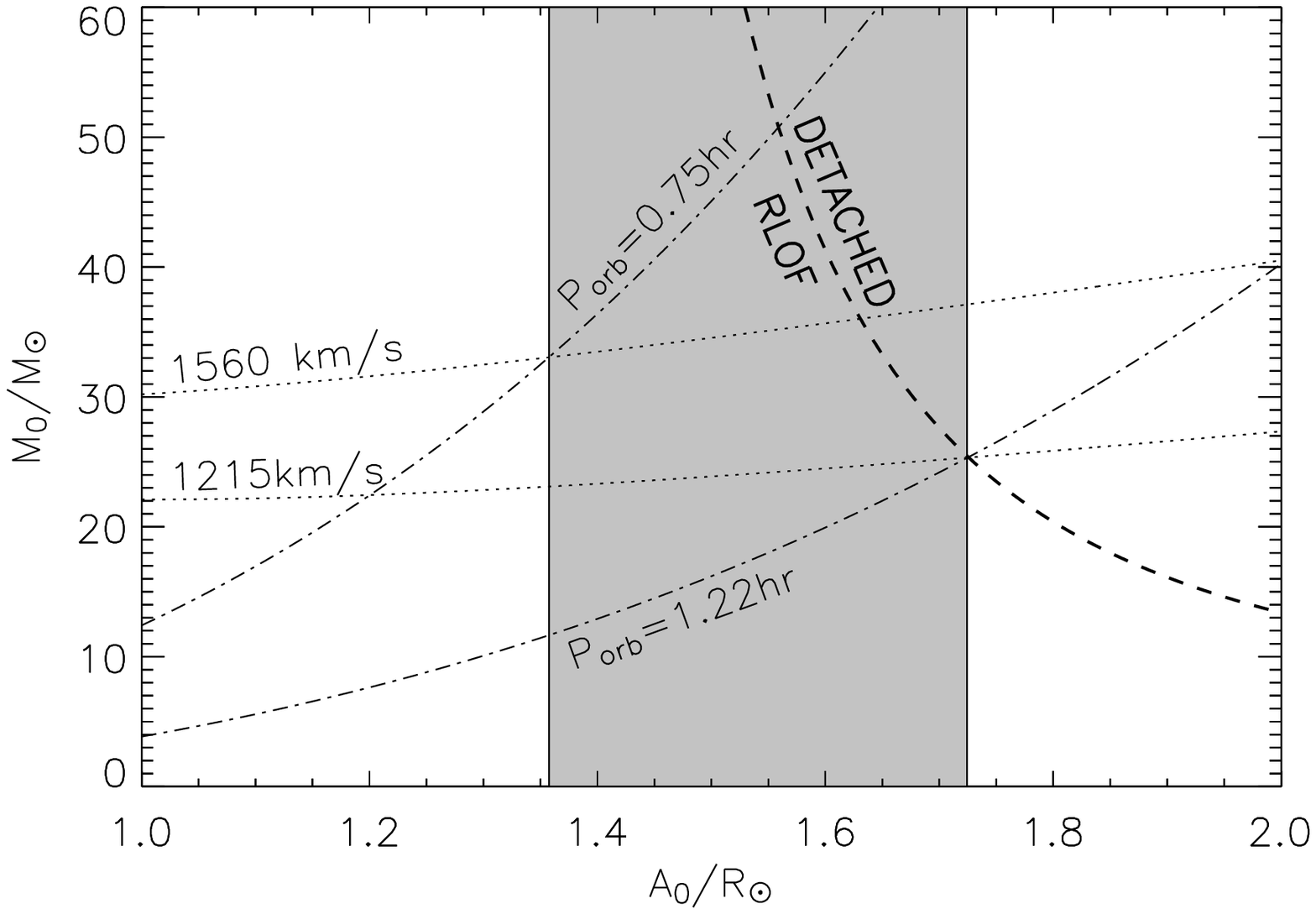}}
\resizebox{8.8cm}{!}{\includegraphics{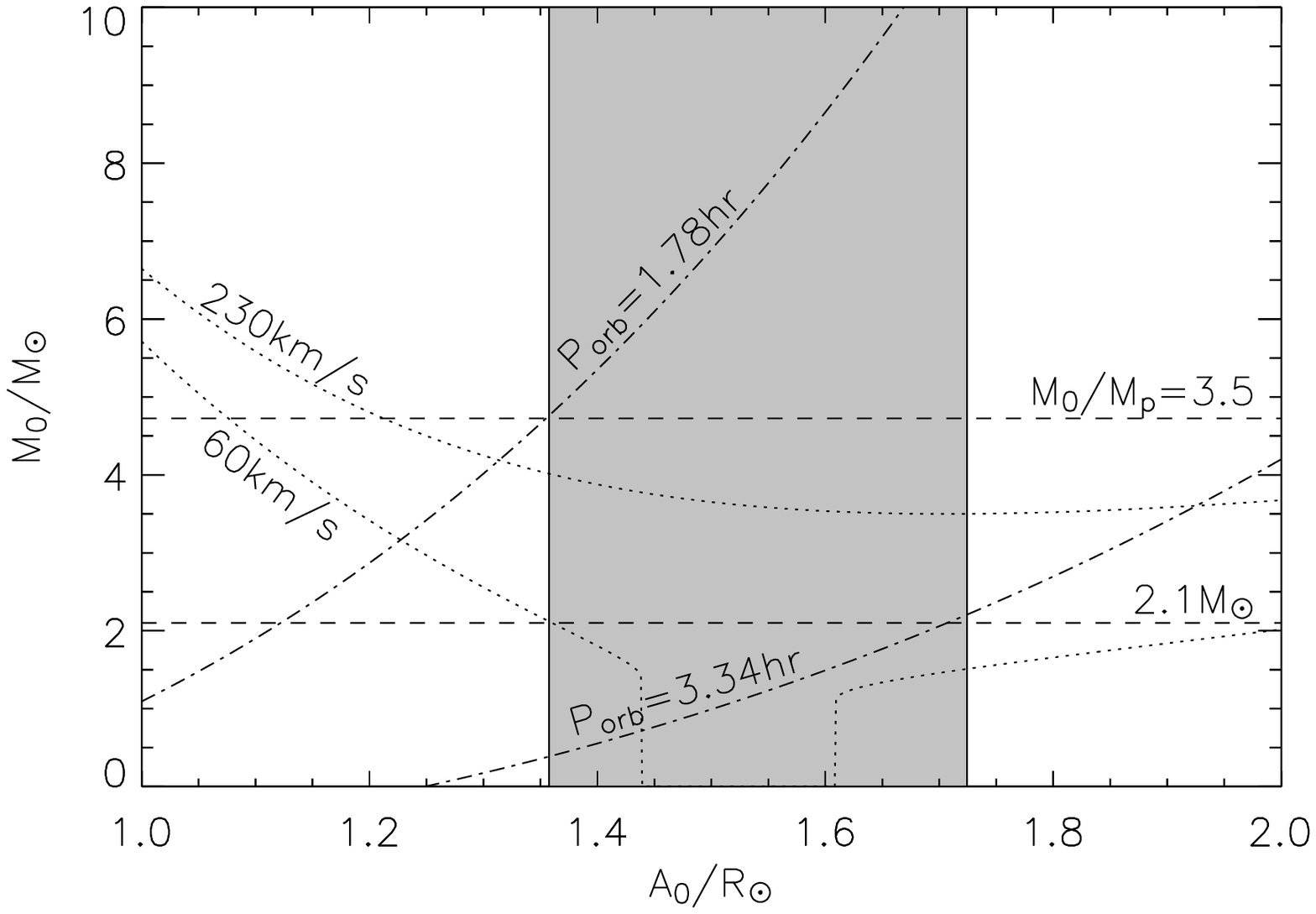}}
\caption{Limits on the pre-SN orbital separation $A_0$ and the helium
star mass $M_0$ for PSR\,J0373-3039 under the assumption that the
last-formed NS was born 100\,Myr ago. The grey-shaded vertical region
represents the band $A(1-e) \le A_0 \le A(1+e)$; the thick
dashed line separates detached from Roche-lobe overflowing systems;
the dotted lines correspond to upper limits on the mass of the helium
star for different kick velocity magnitudes; and the dashed horizontal
lines indicate the mass limits related to the stability of RLOF and
the minimum helium star mass required for the formation of a NS. The
dash-dotted curves represent lines of constant orbital period $P_{\rm
orb}$. In the right-hand panel, we zoom in on the parameter space where
the helium star is overflowing its Roche lobe.}
\label{lim100}
\end{figure*}

The requirements that the post-SN orbit must pass through the position
of the two stars at the time of the explosion and that $\cos^2 \phi
\le 1$, limit the pre-SN orbital separation to the range $A(1-e) \le
A_0 \le A(1+e)$ (Flannery \& van den Heuvel 1975). The range is
independent of the helium star mass and the magnitude of the NS kick,
and is shown by the grey-shaded region in Fig.~\ref{lim100}.

From Eqs.\ (\ref{eq1}) and (\ref{eq2}) it is clear that, for a given
pair of ($M_{0},A_{0}$) there is {\em no unique solution} for the kick
magnitude $V_{k}$ consistent with the post-SN properties (cf.\
DvdH). Instead Fryer \& Kalogera (1997) have shown that 
the absolute requirement $\cos^2 \phi \ge 0$ yields an upper limit for
the mass $M_0$ of the helium star, for every pair of ($A_{0},V_{\rm
k}$) values:
\begin{eqnarray}
M_0 & & \le -M_{\rm p} + k^2 \left( M_{\rm p} + M_{\rm c} 
  \right) \left( A_0/A \right)  \nonumber \\
  & & \! \times \Big\{ \!\! - 2 \left( A/A_0 \right) 
  \left( 1 - e^2 \right)^{1/2} \left[ \left( A/A_0 \right)^2 
  \left( 1 - e^2 \right) - k \right]^{1/2}  \nonumber \\
  & & \! + 2 \left( A/A_0 \right)^2 
  \left( 1 - e^2 \right) - k \Big\}^{-1},  \label{eq3}
\end{eqnarray}
where
\begin{equation}
k = 2\, {A \over A_0} - \left[ {{V_{\rm k}^2\, A} \over 
  {G \left( M_{\rm p} + M_{\rm c} \right)}} + 1 \right].  
  \label{eq4}
\end{equation}
The equality in Eq.~(\ref{eq3}) is valid {\em only} if the kick is
assumed to be restricted in the plane of the pre-SN orbit ($\cos^2
\phi = 0$), and it is only then that the value of $V_{\rm k}$ can be
viewed as an exact solution. For this reason, no strict upper limit on
the magnitude of the kick velocity results from Eq.~(\ref{eq3}), which 
is in contrast to the conclusion obtained by DvdH. As can be seen from
the dotted lines in Fig.~\ref{lim100}, the upper limit on $M_0$
increases with increasing values of $V_{\rm k}$.

%\clearpage

%\clearpage

For a given helium star mass $M_0$, the maximum stellar radius
reached by the second-born NS's direct progenitor sets an additional
divide separating detached from Roche-lobe-filling systems in the
$\left( A_0, M_0 \right)$ parameter space. The divide is represented
by the thick dashed line in the left-hand panel of
Fig.~\ref{lim100}. It follows that in order for the progenitor of
PSR\,J0737-3039 to be detached just before the helium star's SN
explosion, the helium star must be more massive than $\simeq
25\,M_\odot$ and the kick magnitude must be in excess of $\simeq
1200$\,km\,s$^{-1}$. Although such high kick magnitudes have been
discussed in the past (e.g., the guitar nebula - Cordes et al. 1993),
helium stars of such high mass are not very likely at the time of the
SN explosion, given the strong wind mass loss associated with them
(Woosley et al. 1995). In addition, such high-mass helium stars are
expected to end up as a black hole instead of a NS (e.g., Fryer \&
Kalogera 2001; Tauris \& van den Heuvel 2004). Nevertheless, none of
the above are strict constraints, and therefore we conclude that,
although helium stars more massive than $\simeq 25\,M_\odot$ cannot be
excluded, they are most probably highly unlikely. Hence it appears
more reasonable to consider that the helium-star progenitor of the
last-born NS was filling its Roche lobe and was transfering mass onto the
first-born NS at the time of its SN explosion, in agreement with DvdH.

The fate of NS and helium star binaries undergoing mass transfer from
the helium star depends on the orbital period and the mass of the
donor star at the {\it onset of the mass transfer phase}. Since our
analysis yields orbital periods and helium star masses {\it just
before the helium star's SN explosion}, the derivation of exact
constraints in principle requires detailed mass transfer calculations
to map the pre-SN parameter space to the viable parameter space at the
onset of RLOF. However, the details of mass-transfer sequences and the
exact mapping mentioned above depends on the assumptions in
the stellar evolution code adopted (see comparison of results from
three studies of this topic: Dewi \& Pols 2003, Ivanova et al.\ 2003,
and Dewi et al.\ 2002). Instead we can use the qualitative effect of
such a mass transfer phase to derive robust constraints on the NS
progenitor properties. In agreement with other studies Ivanova et al.\
(2003) found that mass transfer from a helium star that is more
massive than the NS companion by a factor greater than 3.5 leads to a
delayed dynamical instability which prevents the formation of a
DNS. Since the pre-SN helium-star mass is bound to be slightly smaller
than the mass at the onset of RLOF, the condition $M_0/M_{\rm p} \le
3.5$ leads to a rather conservative upper limit of $4.7\,M_\odot$ for
the mass of the helium star that formed the companion to
PSR\,J0737-3039. This upper limit is represented by a dashed
horizontal line in Fig.~\ref{lim100}.

A lower limit on the mass $M_0$ of the helium star arises from the
requirement that the helium star must be massive enough to evolve into
a NS instead of a white dwarf. However, the value of this lower limit
depends on the modelling of massive stars as well as on whether or not
the star is affected by binary evolution processes. Current models of
helium star evolution indicate that the minimum mass required to form
a NS ranges from $2.1$ to $2.8\,M_\odot$ (Habets 1985, Tauris \& van
den Heuvel 2004). Here we adopt a conservative value of
$2.1\,M_\odot$. From Fig.~\ref{lim100}, it can be seen that the lower
limit on $M_0$ imposes a lower limit of $60\,{\rm km\, s^{-1}}$ on the
magnitude of the kick velocity imparted to the second NS. Any higher
value for the minimum progenitor mass $M_{0}$ shifts the minimum kick
velocity to higher values (as it is evident from
Fig.~\ref{lim100}). In particular, if the lower limit on $M_0$ would
increase to $2.3\,M_\odot$ as in DvdH, the minimum required kick
velocity is $78\, {\rm km\,s^{-1}}$. The small difference with the
lower limit of $70\, {\rm km\,s^{-1}}$ obtained by DvdH is due to the
different age (and thus different post-SN orbital parameters) adopted
by these authors.

Finally, an upper limit on the magnitude of the kick velocity imparted
to the last-born NS may be derived from the condition that the binary
must remain bound after the SN explosion. The upper limit depends on
the pre-SN helium star mass and orbital separation range, and is given
by $V_{\rm k}/V_{\rm r} = 1 + \sqrt{2 \left( M_{\rm p} + M_{\rm c}
\right) /\left( M_{\rm p} + M_0 \right)}$ (e.g. Brandt \&
Podsiadlowski 1995, Kalogera \& Lorimer 2000). For the mass and
orbital separation constraints ($2.1 \le M_0/M_\odot \le 4.7$ and
$1.36 \le A_0/R_\odot \le 1.72$) derived above, the largest possible
kick velocity is $\simeq 1560$\,km\,s$^{-1}$.

\section{Kick velocity distributions}

For a given kick magnitude $V_{\rm k}$ and a given set of post-SN
orbital parameters $\left( A, e \right)$, Eqs.~(\ref{eq1})
and~(\ref{eq2}) form a set of two algebraic equations relating the
pre-SN orbital separation $A_0$ and the NS progenitor mass $M_0$ to
the polar angle $\theta$ and the azimuthal angle $\phi$ that define the
kick direction with respect to the helium star's pre-SN orbital
velocity. Here we use the constraints on $A_0$ and $M_0$ to derive
constraints on the kick direction that must be satisfied for a given
$V_{\rm k}$ value. It follows that for kick velocities between 60 and
1560\,km\,s$^{-1}$, the polar angle $\theta$ is restricted to the range 
$113^\circ \le \theta \le 180^\circ$, so that the kick is generally
directed opposite to the orbital motion.

We point out that, assuming an isotropic kick distribution, the
constraints on $\theta$ and $\phi$ can be used to derive the
likelihood of the kick magnitude $V_{\rm k}$: the more restricted the
kick direction is, for a given $V_{\rm k}$ value, the lower the
likelihood is. Formally this kick-magnitude likelihood $\Lambda \left(
V_{\rm k} \right)$ is obtained by:
\begin{equation}
\Lambda \left( V_{\rm k} \right) = {1 \over {4\, \pi}} 
\int_{\theta_1}^{\theta_2} \sin \theta\, d\theta
\int_{\phi_1}^{\phi_2} d\phi,  \label{eq5}
\end{equation}
where the boundaries $\theta_1$, $\theta_2$, $\phi_1$, $\phi_2$ of the
admissible region are functions of the kick velocity magnitude $V_{\rm
k}$, and the boundaries $\phi_1$ and $\phi_2$ are usually also
functions of the polar angle $\theta$. Under the assumption that the
kick-velocity magnitude is independent of the direction of the kick,
the probability $P \left( V_{\rm k} \right)$ that the
second-born NS received a kick of magnitude $V_{\rm k}$ is then
obtained by normalising the likelihood so that the integral over all
allowed kick velocities is equal to unity. 

\begin{figure}
\resizebox{8.8cm}{!}{\includegraphics{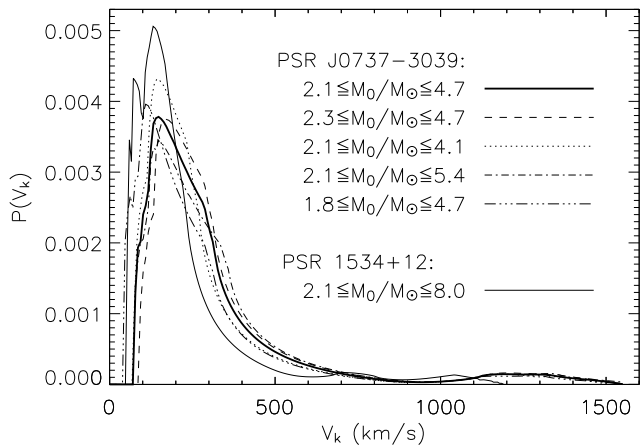}}
\caption{Probability distribution function of kick-velocity magnitudes
$V_{\rm k}$ yielding viable progenitors for PSR\,J0737-3039 for
different mass ranges of the second-born NS's progenitor. 
%The solid line corresponds to the helium star mass range $2.1\,M_\odot
%\le M_0 \le 4.7\,M_\odot$ (see text), the dashed line to $2.3\,M_\odot
%\le M_0 \le 4.7\,M_\odot$, the dotted line to $2.1\,M_\odot \le M_0
%\le 4.1\,M_\odot$, and the dash-dotted line to $2.1\,M_\odot \le M_0
%\le 5.4\,M_\odot$.
The non-zero probabilities for kick velocities larger than $\simeq
1000$\,km\,s$^{-1}$ correspond to kicks for which the majority of the
admissible solutions for $\theta$ and $\phi$ are directed opposite to
the pre-SN orbital velocity and perpendicular to the line connecting
the components' centers of mass. For comparison, the kick-velocity
distribution for PSR\,1534+12 is also shown. The mass range for the
latter is $2.1\le M_0/M_\odot \le 8.0$.}  
\label{pvk}
\end{figure}

The probability distribution function $P \left( V_{\rm k} \right)$ for
PSR\,J0737-3039 is plotted in Fig.~\ref{pvk} (thick solid line). The
curve has a clear maximum at $\simeq 150$\,km\,s$^{-1}$ which
represents the most probable kick magnitude imparted to the pulsar
companion at birth. In order to assess the sensitivity of the
distribution to our helium-star mass constraints, we also show curves
corresponding to different lower and upper limits on the mass of the
helium star progenitor of the second-born NS. As can be seen from the
figure, the distribution is not very sensitive to changes in the upper
limit on the allowed helium star mass range. In the particular case of
a slightly higher lower limit of $2.3\,M_\odot$ on the mass of the
helium star, the peak in the distribution shifts to $\simeq
165$\,km\,s$^{-1}$. For comparison, the kick-velocity distribution for
PSR\,1534+12 is also shown in Fig.~\ref{pvk}.

%\clearpage

%\clearpage

\section{Discussion}

We derived constraints on the pre-SN progenitor of the newly
discovered relativistic binary pulsar PSR\,J0737-3039. For an assumed
age of 100\,Myr, the tight limits on the pre-SN orbital separation
($1.36 \le A_0/R_\odot \le 1.72$) imply that the progenitor consists
of the first-formed NS in orbit around a helium star (and not its
hydrogen-rich progenitor since the system would then be in a
common-envelope phase with a spiral-in time scale that is much shorter
than the evolutionary time scale leading to the SN explosion). We
found that the helium star is most likely overflowing its Roche lobe
and constrained its mass to be between $2.1\,M_\odot$ and
$4.7\,M_\odot$. The lower limit of $2.1\,M_\odot$ implies that a birth
kick with a velocity of at least 60\,km\,s$^{-1}$ was imparted to the
second-born NS, in agreement with the minimum kick velocity derived by
DvdH.  From the condition that the binary must remain bound after the
second SN explosion, we derived an upper limit for the
kick velocity of 1560\,km\,s$^{-1}$. This is in contrast to the upper
limit of 230\,km\,s$^{-1}$ derived by DvdH, which is valid only if the
kick is restricted to the pre-SN orbital plane. These results are
fairly insensitive to the adopted age: if the system were only 50\,Myr
old, the progenitor and kick constraints are $1.27\,R_\odot \le A_0
\le 1.57\,R_\odot$ and $65\, {\rm km\,s^{-1}} \le V_{\rm k} \le 1610\, 
{\rm km\,s^{-1}}$. The allowed helium star mass range is independent
of the adopted age. 

We furthermore extended the constraints on NS formation and, for the
first time, derived a probability distribution for the kick magnitude
imparted to the second-born NS in a DNS binary (PSR J0737-3039). The
distribution exhibits a clear maximum at $150$\,km\,s$^{-1}$ which is
fairly independent of the allowed helium star mass range and the
assumed age of the system. In addition, a small secondary peak was
found for kick velocities larger than $\simeq 1000$\,km\,s$^{-1}$
which mainly correspond to kicks directed opposite to the pre-SN
orbital velocity and perpendicular to the line connecting the
components' centers of mass.

We also applied the analysis described above to the other two
relativistic DNS systems known in the galactic disk. These systems may
arise from detached as well as semi-detached pre-SN progenitors. An
upper limit for the mass of the helium star in these progenitors is 
therefore given
by the largest possible helium star mass forming a NS instead of a
black hole. If we set this upper limit at $8\,M_\odot$, the most
likely kick velocity imparted to the second-born NS in PSR\,1913+16 is
240\,km\,s$^{-1}$. In addition, it turns out that kicks smaller than
$170$\,km\,s$^{-1}$ are allowed but have a vanishingly small
probability. This is in contrast to the findings of Fryer \& Kalogera
(1997) and Dewi \& Pols (2003) who found minimum kick velocities of
260\,km\,s$^{-1}$ and 70\,km\,s$^{-1}$, respectively. Note, however,
that in the derivation of the kick-velocity distribution for
PSR\,1913+16 we did not yet take into account the measured space
velocity as was done by Wex et al. (2000). We will include this in a
forthcoming investigation on the spin-orbit misalignment of
PSR\,J0737-3039A, where we will also present a more detailed 
comparison between the possible kick velocities and kick directions
imparted to the last-born NS in PSR\,1913+16 and PSR\,J0737-3039.
Finally, for PSR\,1534+12, we find that the most likely kick velocity
imparted to the second-born NS is 130\,km\,s$^{-1}$ and that kick
velocities below $100$\,km\,s$^{-1}$ have a vanishingly small
probability. 

\acknowledgments 
This work is partially supported by a NSF Gravitational Physics grant,
a David and Lucile Packard Foundation Fellowship in Science and
Engineering grant, and NASA ATP grant NAG5-13236 to VK.

\end{document}